\documentclass[review,12pt]{elsarticle}

\journal{Computational Materials Science}

\usepackage[british]{babel}
\usepackage{hyphenat}

\usepackage{lineno,hyperref}
\modulolinenumbers[5]

\usepackage{graphicx,amssymb}
\usepackage{color}

\begin{document}

\begin{frontmatter}


\title{Ab initio molecular dynamics simulations of negative thermal expansion in ScF$_3$: The effect of the supercell size}


\author[ISSP]{D.\ Bocharov}
\ead{bocharov@latnet.lv}
\cortext[cor1]{Corresponding author}
\ead[url]{http://www.cfi.lu.lv/}
\author[PSI]{M.\ Krack}
\author[ISSP]{Yu.\ Rafalskij}
\author[ISSP]{A.\ Kuzmin}
\author[ISSP]{J.\ Purans}

\address[ISSP]{Institute of Solid State Physics, University of Latvia,
Kengaraga Street 8, LV-1063 Riga, Latvia}

 \address[PSI]{Paul Scherrer Institute, Forschungsstrasse 111, CH-5232 Villigen PSI, Switzerland}


\begin{abstract}
Scandium fluoride (ScF$_3$) belongs to a class of negative thermal expansion (NTE) materials. It shows a strong lattice contraction up to about 1000 K 
switching to expansion at higher temperatures. Here the NTE effect in ScF$_3$ is studied in the temperature range from 300 K to 1600 K 
using ab initio molecular dynamics (AIMD) simulations in the isothermal-isobaric (NpT) ensemble. 
The temperature dependence of the lattice constant, inter-atomic Sc--F--Sc bond angle distributions and the Sc--F and Sc--Sc radial distribution functions 
is obtained as a function of supercell size from $2a \times 2a \times 2a$ to $5a \times 5a \times 5a$ where $a$ is the lattice parameter of ScF$_3$.
 A comparison with the experimental Sc K-edge EXAFS data at 600 K is used to validate the accuracy of the AIMD simulations.
Our results suggest that the AIMD calculations are able to reproduce qualitatively the NTE effect in ScF$_3$, 
however a supercell size larger than $2a \times 2a \times 2a$
should be used to account accurately for dynamic disorder. 
The origin of the NTE in ScF$_3$ is explained by the interplay between expansion and rotation of ScF$_6$ octahedra.
\end{abstract}

\end{frontmatter}


 \newpage

\section{Introduction}

Materials with negative thermal expansion (NTE), contracting upon heating,
are not only of great interest from a fundamental physics point of view, but have also a high industrial importance \cite{Evans1997,Takenaka2012,Dove2016}.
Composites containing NTE components can possess zero thermal expansion, making them suitable 
for high-precision devices, like space telescope mirrors, teeth fillings, substrates in microelectronics, fuel cells, thermoelectric converters and for many other applications \cite{BookNTE}.  

During the last decade, fluorides of various metals, including ScF$_3$, have attracted attention as a new class of materials with NTE \cite{WANG2015106}. ScF$_3$ is a peculiar compound, which has simple ReO$_3$-type cubic structure and surprisingly strong NTE effect appearing as a decrease of the lattice constant over a wide range of temperatures from 10 K  to 1100 K, while the positive expansion of the lattice occurs at higher temperatures \cite{JACS}. 
This makes ScF$_3$ an excellent study subject for a deeper understanding of the NTE phenomenon.  Note that the NTE of pure ScF$_3$ can be affected by reducing crystallites size  \cite{Yang2016,Hu2018} or by substituting  
the scandium atoms with yttrium \cite{Morelock2013}, titanium \cite{Morelock2014}, iron \cite{Hu2014,Han2016,Chen2017}, gallium \cite{Hu2014} or aluminium \cite{Han2016,Morelock2015} atoms. 

The NTE effect is often explained based on the vibrational mechanism in terms of the 
so-called rigid unit modes (RUMs) model \cite{Dove2016,Pryde1997,Tao2003}, 
which involves coupled vibrations of the ScF$_6$ octahedra. 
When two neighbouring rigid ScF$_6$ octahedra librate in opposite directions, the distance between the scandium atoms located at their centres decreases leading to the lattice contraction.

The rigidity of ScF$_6$ octahedra and coupling of their relative motion are determined by the
strength of the Sc--F chemical bonding. Therefore, the accurate description of the NTE effect should account for the interaction between the lattice, phonons and electrons. 
This challenging problem can be addressed using the method of molecular dynamics (MD), which provides a natural way to include thermal disorder in simulations \cite{Mladenovic2018}.
Moreover, the use of ab initio molecular dynamics (AIMD), being computationally much more demanding than classical MD \cite{Bocharov2019},  allows one to account explicitly for chemical bonding and its anisotropy, which are most likely important for the interpretation of the NTE effect. 
 
Until now the AIMD method was used to study the NTE of ScF$_3$ in
\cite{neutrons,Lazar2015}. 
Both works employed ab initio Born-Oppenheimer molecular dynamics implemented in 
the VASP code \cite{VASP} based on a plane wave basis set.  

In \cite{neutrons} the AIMD simulations were performed to study anharmonic effects in the temperature range between 7 K and 750 K for a $3a \times 3a \times 3a$ ($a$ is the lattice parameter) supercell 
containing 108 atoms. Based on the analysis of the MD trajectories, it was concluded that the motion of fluorine atoms is largely uncorrelated and 
strongly anisotropic in the direction orthogonal to the Sc--F--Sc bonds.
No attempt was made by Li et al.\ \cite{neutrons} to reproduce the NTE behaviour of ScF$_3$. 

The NTE effect was studied using AIMD simulations by Lazar et al. \cite{Lazar2015} employing the isothermal-isobaric (NpT) ensemble with a small $2a \times 2a \times 2a$ supercell containing only 32 atoms. The simulations were performed in the 
temperature range between 200 K and 1400 K. The experimental behaviour of the lattice constant $a$ was reproduced after its normalization relative to the value calculated at 200 K. These simulations also predicted that the ScF$_6$ octahedra remain 
non-distorted at all temperatures. The NTE behaviour of ScF$_3$ was explained by an interplay between the linear thermal expansion of the Sc--F bonds  and a decrease of the average Sc--F--Sc bond angles due to octahedra tilting motion \cite{Lazar2015}.

Though the AIMD method is a powerful tool to describe the NTE in ScF$_3$, 
its accuracy is limited by several issues. 

It was demonstrated recently \cite{Oba2019}, that the AIMD simulations based on the Newton’s equations of motion underestimate the magnitude of the
NTE in the entire temperature range due to a neglect of the zero point (quantum) atomic motion.
The problem is particularly evident at low temperatures below 500 K \cite{Oba2019}.

Another issue, which is the topic of the present study, is related to the size of the supercell used in the simulations. 
To describe the librational  motion of ScF$_6$ octahedra, one needs at least eight octahedra placed in a simulation box consisting of 2$\times$2$\times$2 multiples of the primitive cubic cell of ScF$_3$ \cite{Lazar2015}. However, one can expect that such small supercell will strongly influence the lattice dynamics, in particular, long wavelength phonons, and, as a result, the correlation effects in the atomic motion will be overestimated. Since AIMD simulations are computationally expensive,
the choice of the supercell size is critical. 

Note that the results of AIMD simulations can be validated by direct comparison with the results provided by experimental methods sensitive to both average structure and disorder 
such as the pair distribution function (PDF) analysis \cite{Keen2015}  or 
the extended X-ray absorption fine structure (EXAFS) spectroscopy \cite{Kuzmin2014IUCR}.
EXAFS, being also sensitive to high-order correlation functions, provides a unique possibility to probe the local structure and lattice dynamics of materials. 
It was  successfully used by us previously to validate MD simulations for different materials such as SrTiO$_3$ \cite{kuzmin2009quantum}, ReO$_3$ \cite{Kalinko2009reo3}, ZnO \cite{Kuzmin2016zpc},  UO$_2$ \cite{Bocharov2016_UO2} and Cu$_3$N \cite{Cu3N}. 

In this study, we employ AIMD simulations to reproduce the NTE effect in ScF$_3$ and investigated the influence of the supercell size on the local structure and dynamics. 
In particular, we will demonstrate that the use of the 
smallest supercell ($2a \times 2a \times 2a$) fails to describe thermal disorder accurately leading to a broadening of the inter-octahedral Sc--F--Sc bond angle distribution and of the peaks in the radial distribution function starting from the third coordination shell of scandium. At the same time, the details of the lattice dynamics due to the NTE effect  
are well reproduced for larger supercells, starting from $4a \times 4a \times 4a$.
The explanation of the NTE effect in ScF$_3$ is discussed based on the AIMD results.

\begin{figure}[t]
	\begin{center}
		\includegraphics[width=0.7\linewidth]{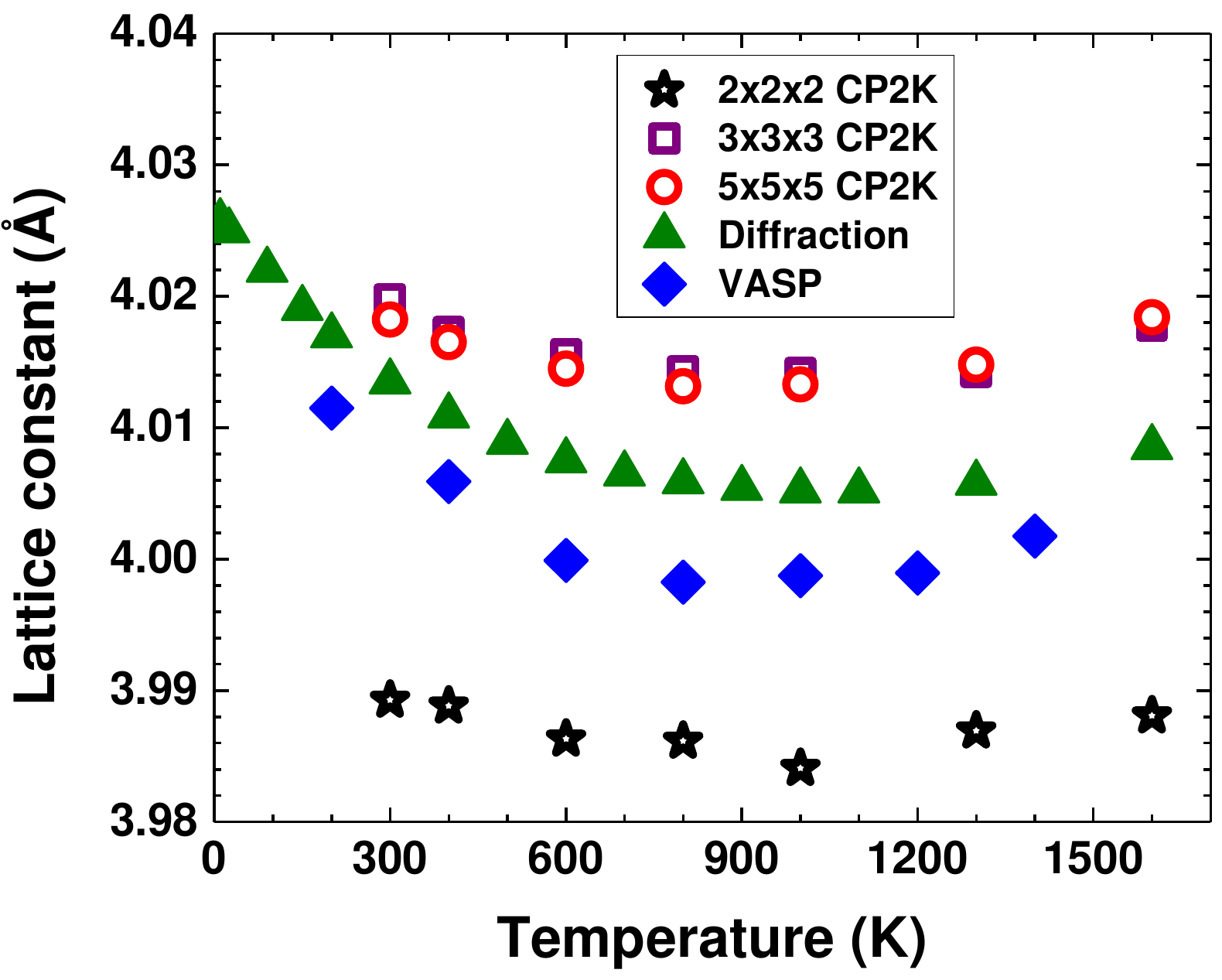}
	\end{center}
	\caption{Comparison of the temperature dependences of the lattice constant of ScF$_3$ calculated in this work for $2a \times 2a \times 2a$ (open asterisks), $3a \times 3a \times 3a$ (open squares), $5a \times 5a \times 5a$ (open circles) 
		supercell using CP2K code, calculated in \protect\cite{Lazar2015} for $2a \times 2a \times 2a$ supercell by VASP code (solid diamonds) and 
		experimental data from \protect\cite{JACS} (solid triangles).}
	\label{fig1}
\end{figure}

\begin{figure}[t]
	\begin{center}
		\includegraphics[width=1\linewidth]{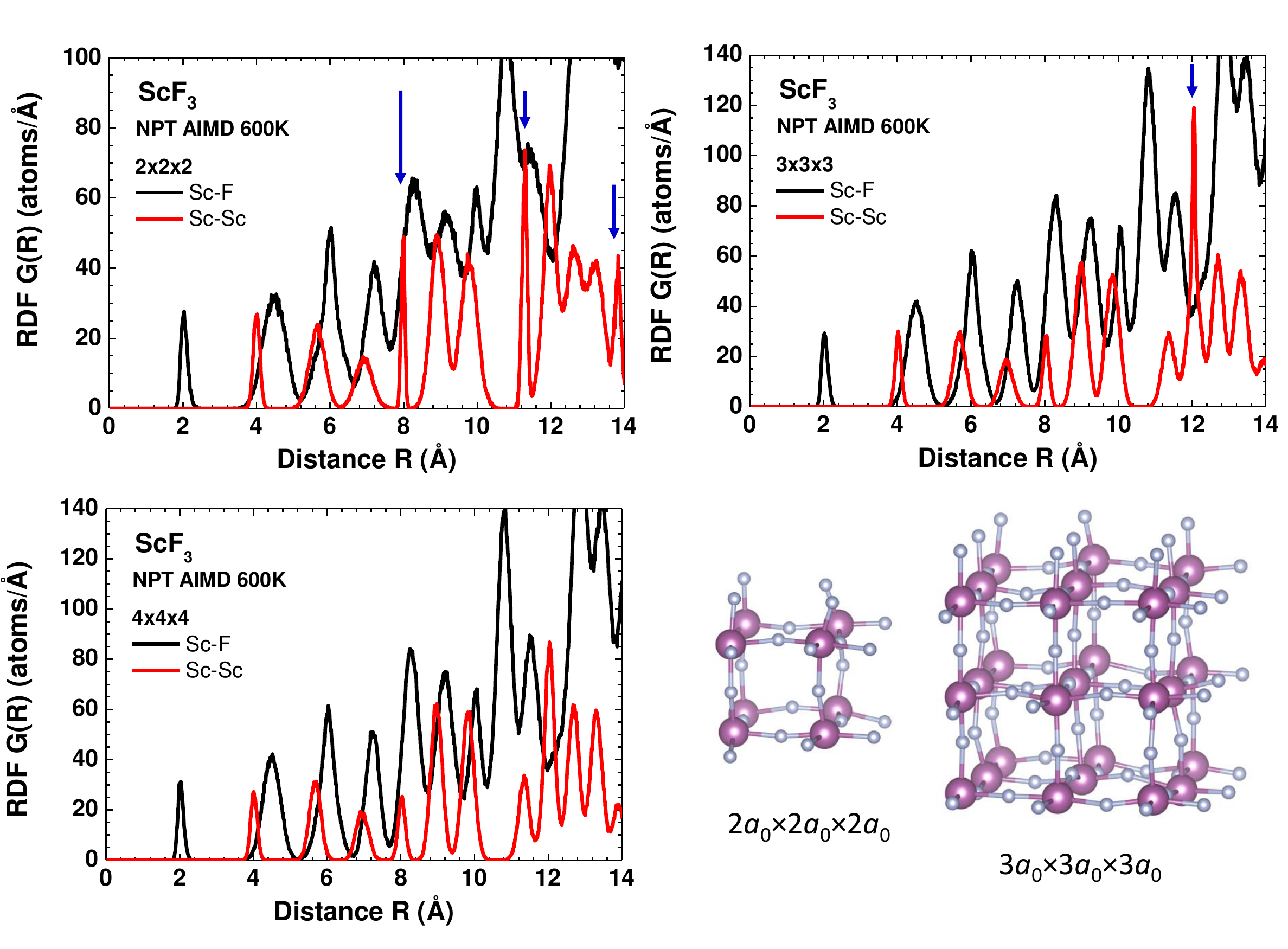}
	\end{center}
	\caption{Radial distribution functions (RDFs) $G$($R$) for the Sc--F and Sc--Sc atom pairs at $T$ = 600 K calculated from the AIMD simulations for $2a \times 2a \times 2a$,  $3a \times 3a \times 3a$  and $4a \times 4a \times 4a$  supercell sizes. The models of the $2a \times 2a \times 2a$ and $3a \times 3a \times 3a$  supercells are also shown. }
	\label{fig2}
\end{figure}

\begin{figure}[t]
	\begin{center}
		\includegraphics[width=0.8\linewidth]{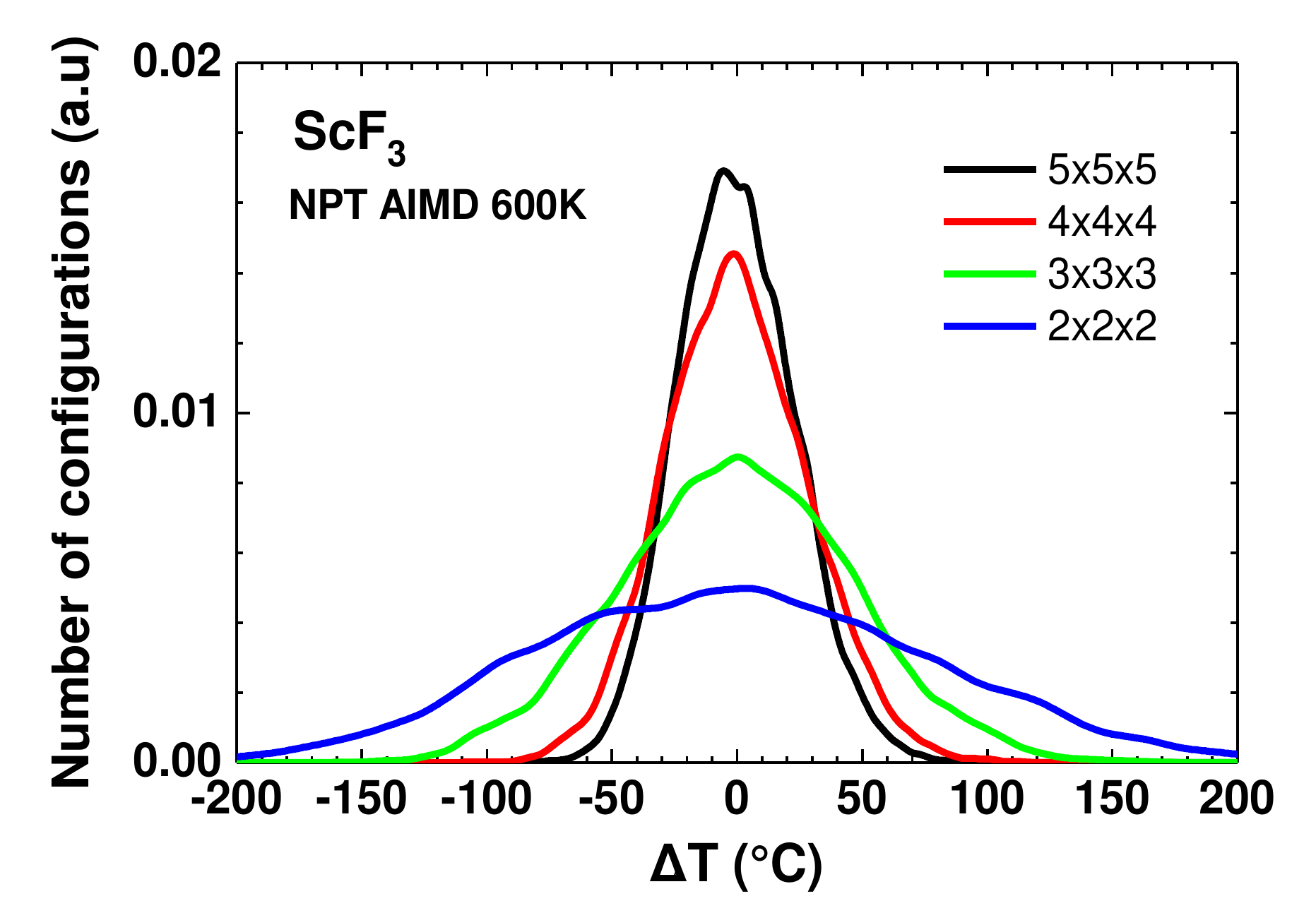}
	\end{center}
	\caption{Instantaneous distribution of temperature for supercells of different sizes at $T$ = 600 K. }
	\label{fig3}
\end{figure}

\section{Ab initio molecular dynamics}
\label{aimd}

Understanding of the NTE effect in ScF$_3$ requires detailed and accurate knowledge 
of its temperature-dependent structure and lattice dynamics. 
This information is obtained in the present study using the AIMD simulations.

Our simulations were based on Kohn--Sham density functional theory (DFT)
\cite{Kohn-PR140pA1133} and were performed in the isothermal–isobaric (NpT) ensemble
at seven different temperatures (300 K, 400 K, 600 K, 800 K, 1000 K, 1300 K and
1600 K) using the CP2K code \cite{Quickstep}. The CP2K code employs a
localized basis set of Gaussian-type orbital functions for the description of
the Kohn-Sham matrix within the framework of the Gaussian Plane Waves method
\cite{Quickstep,GPW}. A cutoff of 600 Rydberg is used for the auxiliary basis set
of plane waves to expand the electronic density. Sc and F atoms are described
by scalar-relativistic norm-conserving Goedecker-Teter-Hutter pseudopotentials
\cite{GTH,Hartwigsen1998,Krack2005} are employed for including 11 ($\lbrack{\rm
Ne}\rbrack\,3s^2\,3p^6\,4s^2\,3d^1$) and 7 ($\lbrack{\rm
He}\rbrack\,2s^2\,2p^5$) valence electrons, respectively. Calculation were
performed at the Gamma point only using MOLOPT basis sets \cite{MOLOPT}
optimized for these pseudopotentials. By performing DFT calculations at $T$ = 0 K 
we found that the PBEsol functional gives the value of the lattice constant
($a$ = 4.027 \AA) closer to the experimental one ($a$ = 4.026 \AA\ \cite{JACS})
than the PBE functional  \cite{PBE}  ($a$ = 4.065 \AA), which  can be important
for proper description of the NTE effect. Therefore, the PBEsol
exchange-correlation functional \cite{PBEsol} was used in all AIMD calculations.

ScF$_3$  model system of increasing size ranging from
$2a \times 2a \times 2a$ to $5a \times 5a \times 5a$ primitive unit
cells containing 32--500 atoms (8--125 unit cells with 4 atoms per unit cell)
were employed in the AIMD simulations.  After  thermalisation MD run of about
15 ps, an AIMD production run of 50 ps was performed for the $3a \times 3a \times 3a$ to $5a \times 5a \times 5a$ supercells 
and 100 ps for the $2a \times 2a \times 2a$ supercell. The required temperature
was maintained during the sampling run using the CSVR (Canonical Sampling
through Velocity Rescaling) thermostat of Bussi et al.\ \cite{Bussi2007}.
The MD time step of 0.5 fs was used in all simulations. 
The values of the lattice parameter $a$ are reported in Fig.~\ref{fig1}: they were calculated by averaging over all atomic configurations obtained during the production run.

The obtained from MD simulations sets of atomic coordinates were used to calculate the radial
distribution functions (RDFs) $G$($R$)  for the Sc--F and Sc--Sc atomic pairs
and are reported in  Fig.~\ref{fig2}. The RDF $G$($R$) is defined as a number of atoms (Sc or F)
located within a distance of $R$ and $R+dR$ ($dR$ = 0.01 \AA) away from the scandium atom. 

Fig.~\ref{fig3} shows instantaneous distribution of temperatures during the
production run for supercells of different sizes at $T$ = 600 K. The temperature
fluctuation is in the range of about $\pm$100 K from the target one for two
largest ($4a \times 4a \times 4a$ and $5a \times 5a \times 5a$)
supercells.  The temperature fluctuation range becomes broader for the
$3a \times 3a \times 3a$ supercell exceeding 100 K and is twice larger for
the  $2a \times 2a \times 2a$ supercell.
The growing broading of the temperature distribution for shrinking
supercells is caused by the decreasing number of degrees of freedom (DOF) for
smaller model systems containing less atoms. A massive thermostatting by
employing a thermostat for each DOF was only applied during the thermalisation
period whereas just one global thermostat was coupled to the actual model system
during the sampling period in order to minimise any bias from the interaction with
the thermostat. Fig.~\ref{fig3} shows that a narrow temperature distribution can be obtained
even with a mild thermostatting if a sufficiently large model system is chosen
which will eventually ensure physically more meaningful results compared to a
small model system requiring massive thermostatting.

\begin{figure}[t]
	\begin{center}
		\includegraphics[width=0.65\linewidth]{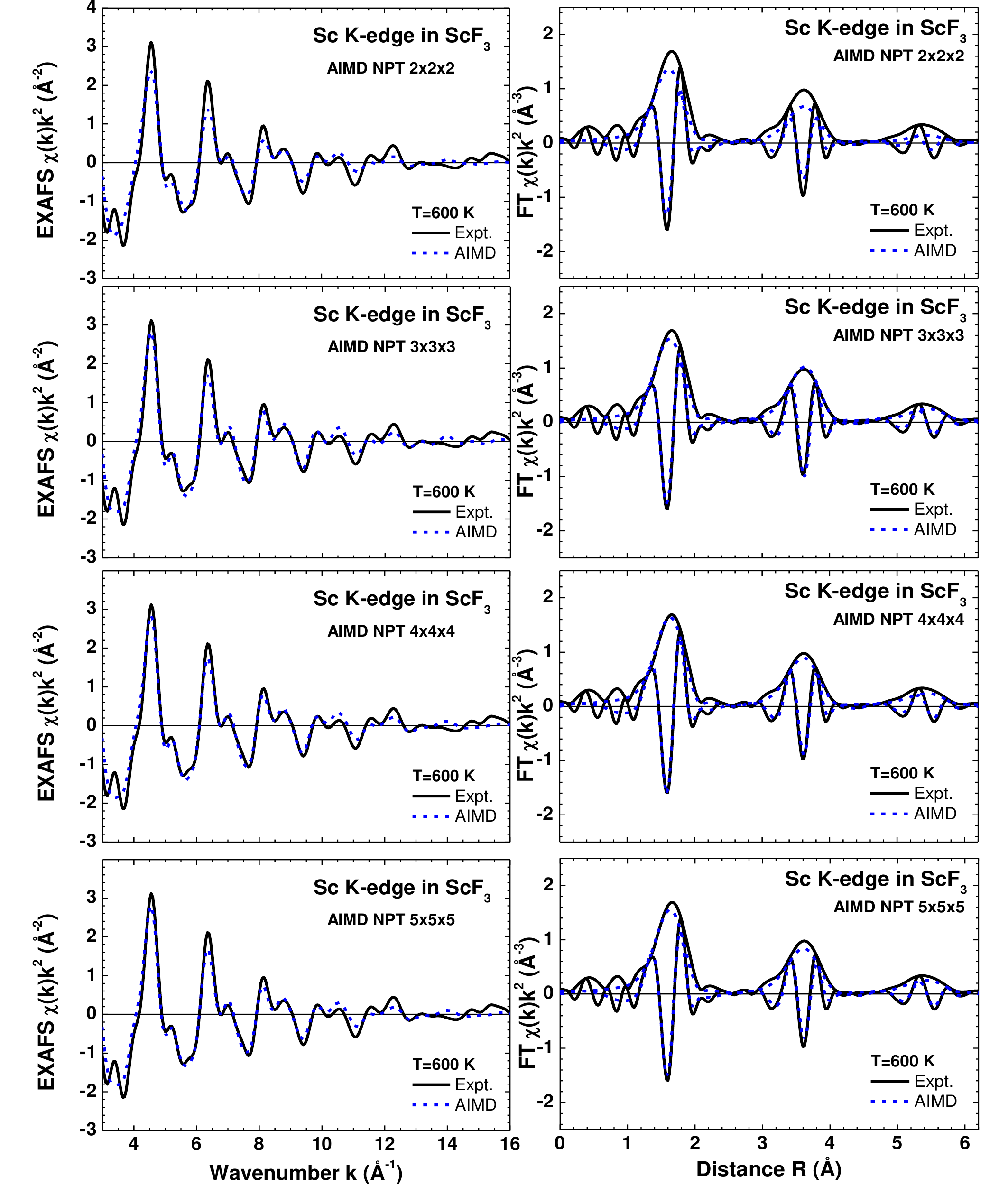}
	\end{center}
	\caption{Experimental (black solid line) and calculated (blue dashed line) Sc K-edge EXAFS $\chi(k)k^2$ and their Fourier transforms (modulus and imaginary parts are presented) at 
		$T$ = 600 K for different supercell sizes.}
	\label{fig4}
\end{figure}

\section{Validation of AIMD simulations}\label{mdexafs}

Temperature-dependent AIMD simulations provide information on the average structure, reported in Fig.~\ref{fig1} in terms of the ScF$_3$ lattice parameter, and on the dynamical (time-dependent) structure, which can be described by distribution functions.
The latter can be experimentally probed by X-ray or neutron total scattering experiments \cite{Hu2018,Hu2016} or by EXAFS spectroscopy \cite{Bocharov2016}. 

The results of AIMD simulations were validated in the present study using the experimental Sc K-edge EXAFS data from \cite{Bocharov2016,Purans2016}, obtained at $T$ = 600 K to minimize the influence of zero point quantum effects \cite{Oba2019}.
Sets of atomic configurations obtained in AIMD NpT simulations               
were used to calculate configuration-averaged EXAFS spectra $\chi(k)$ ($k$ is the photoelectron wavenumber) following the approach described previously \cite{kuzmin2009quantum,Kuzmin2016zpc}.

The Sc K-edge EXAFS spectrum for each configuration was calculated using the
real-space multiple-scattering  FEFF9.64 code \cite{FEFF9,Rehr2000}.
First, the scattering potential and partial phase shifts of Sc and F atoms were obtained within the muffin-tin (MT)
approximation (15\% overlap of the nearest MT-spheres, $R_{\rm MT}$(Sc)=1.31 \AA\ and $R_{\rm MT}$(F)=1.01 \AA) for the cluster of 8.0 \AA\ radius, constructed 
using the crystallographic ScF$_3$ structure \cite{JACS} and centered at the absorbing Sc atom. 
The cluster potential was fixed during all simulations, thus we neglected its small variations due to thermal vibrations. The multiple-scattering contributions were accounted up to the 6th order to guarantee the convergence of the total EXAFS in the $k$-space range of interest.  The photoelectron inelastic losses were accounted within the one-plasmon approximation using the complex exchange-correlation Hedin-Lundqvist potential \cite{Hedin1971}.
The value of the amplitude reduction factor $S_0^2$ was set to 1.0 \cite{Kuzmin2014IUCR,Rehr2000}.

The configuration-averaged Sc $K$-edge EXAFS $\chi(k)k^2$ spectra of ScF$_3$ and their Fourier transforms (FTs) at $T$ = 600 K are shown  in Fig.~\ref{fig4} for several supercell sizes. The Fourier transforms
were calculated using the 10\% Gaussian window function and were not corrected for the backscattering phase shift of atoms, therefore the positions of all peaks are displaced to smaller distances relative to their crystallographic values.
The significantly worse agreement between the experimental and calculated
spectra for the $2a \times 2a \times 2a$ supercell is obvious.

\begin{figure}[t]
	\begin{center}
		\includegraphics[width=0.5\linewidth]{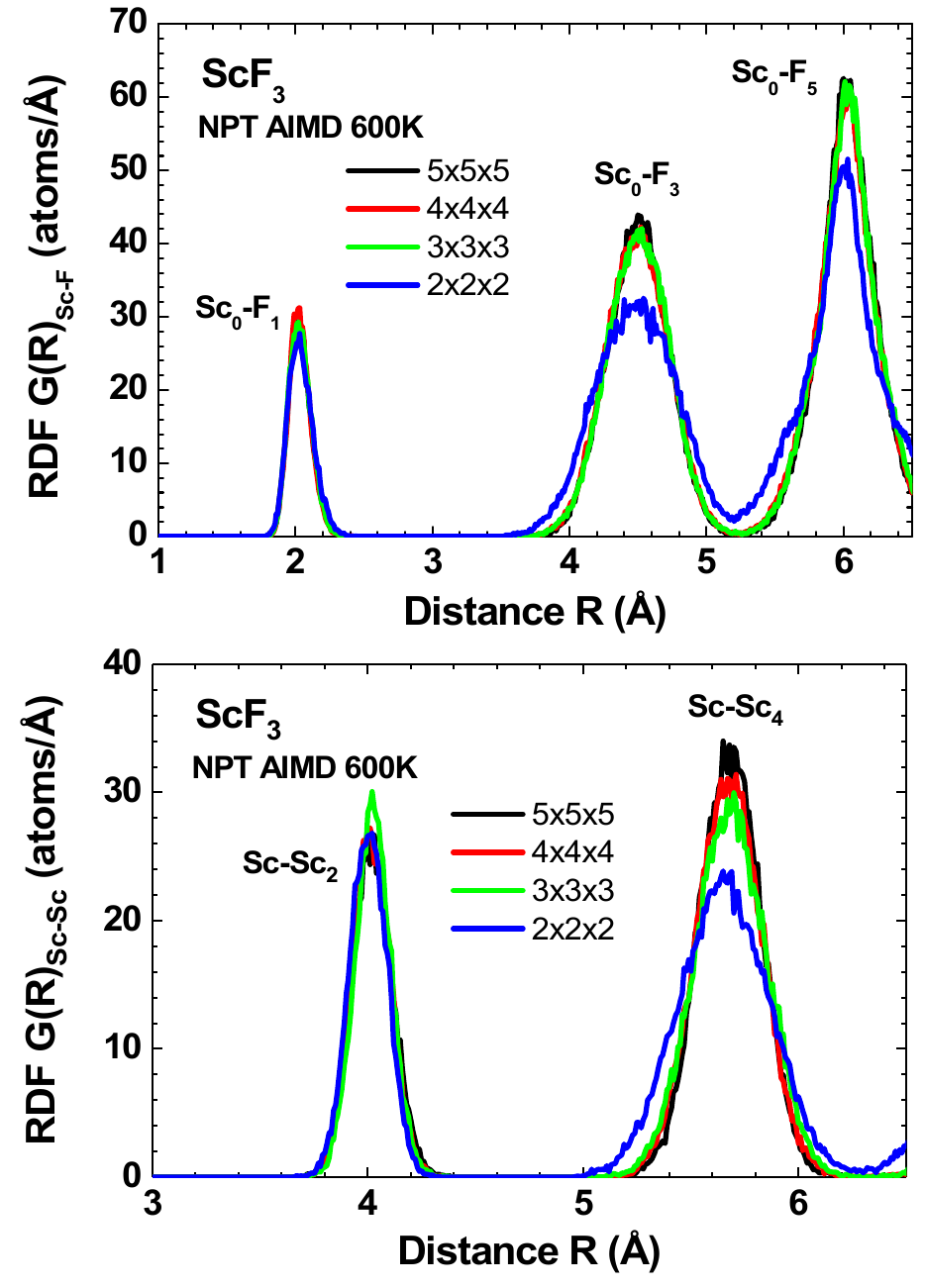}
	\end{center}
	\caption{Radial distribution functions (RDFs) $G$($R$) for the Sc--F  and Sc--Sc  atom pairs at $T$ = 600 K calculated from the AIMD simulations for different supercell sizes. }
	\label{fig5}
\end{figure}

\section{Discussion}\label{disc}

The temperature dependence of the lattice constant of ScF$_3$, determined
from NpT AIMD simulations, are compared with the experimental data obtained by
diffraction measurements  in Fig.~\ref{fig1}. The AIMD simulations
performed in this study for $2a \times 2a \times 2a$, $3a \times 3a \times 3a$ 
and  $5a \times 5a \times 5a$ supercells and by Lazar et al.\ for a $2a \times 2a \times 2a$ supercell
reproduce qualitatively the negative thermal expansion effect
in scandium fluoride up to 1000 K as well as positive expansion at higher
temperatures. The quantitative values of the calculated lattice constant $a$ are close to the experimental diffraction values \cite{JACS} with a deviation being less than $\pm$0.03 \AA. Comparing our results for different supercell sizes, one can conclude that
a reduction of the absolute lattice constant occurs in small $2a \times 2a \times 2a$ supercell. This effect is caused by the overestimated amplitude of the librational  motion of ScF$_6$ octahedra and is clearly observed in the Sc--F--Sc bond angle distribution function (BADF) discussed below. 

Further we will discuss the results obtained at $T$ = 600 K as a representative case. 

Radial distribution functions (RDFs) $G(R)$ for the Sc--F and Sc--Sc atom pairs at $T$ = 600 K  calculated  in the range of $R$ = 0--14 \AA\ for the supercell sizes from   $2a \times 2a \times 2a$ to $4a \times 4a \times 4a$ are shown in Fig.~\ref{fig2}. As one can see, for the smallest supercell sizes, sharp peaks in the Sc--Sc RDF (indicated by arrows in Fig.~\ref{fig2}) are observed  due to periodic boundary conditions (PBC) employed in the simulations. For the $2a \times 2a \times 2a$ supercell with a size of $2a$ = 8.054 \AA,  the three peaks
are due to PBC along $<$1 0 0$>$, $<$1 1 0$>$ and $<$1 1 1$>$ crystallographic directions. For the $3a \times 3a \times 3a$ supercell with a size of $3a$ = 12.081 \AA, 
only one sharp peak due to the PBC along $<$1 0 0$>$ is observed in the $R$-range till 14 \AA. Similar effects should appear also in the Sc--F RDFs, however, they are much less visible 
due to an overlap between closely located shells.   

An enlarged view of the RDFs in the range  of the first five coordination shells of scandium is shown in Fig.~\ref{fig5}. The reduction of the supercell size clearly leads to an increase  
of the peak widths and to a distortion of the peak shape for the smallest supercell. At the same time, the asymmetric shape of the first (Sc--F$_1$) and second (Sc--Sc$_2$) shell 
peaks is close for all supercells, explaining the success of the  $2a \times 2a \times 2a$ supercell model in a description of the NTE lattice behaviour in \cite{Lazar2015}.

\begin{figure}[t]
	\begin{center}
		\includegraphics[width=0.6\linewidth]{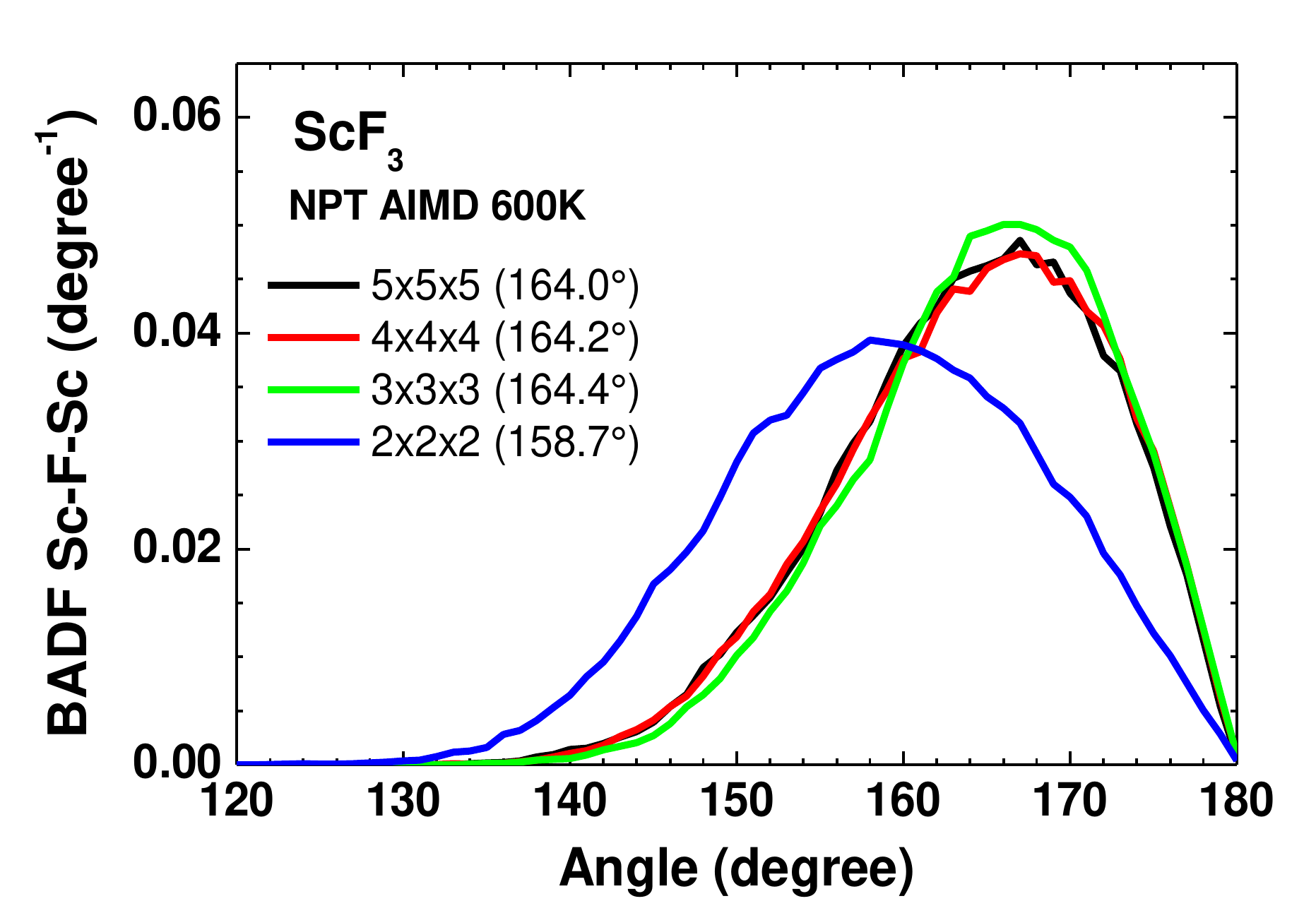}
	\end{center}
	\caption{The inter-octahedral Sc--F--Sc bond angle distribution function (BADF) at $T$ = 600 K for NpT ensemble and $2a \times 2a \times 2a$ to $5a \times 5a \times 5a$ supercell sizes.
	The average values of the Sc--F--Sc bond angle for each BADF are given in brackets.}
	\label{fig6}
\end{figure}

The atomic coordinates obtained in the AIMD simulations were used to calculate the inter-octahedral Sc--F--Sc BADF and the average value of the Sc--F--Sc angle, which are reported in Fig.~\ref{fig6}. Note that in cubic ScF$_3$ the crystallographic angle between the average positions of atoms in the Sc--F--Sc chains is equal to 180$^\circ$.
However, the value of the average Sc--F--Sc angle will be always smaller when calculated from mean distances due to vibrations of fluorine atoms perpendicular to the direction of the Sc–F–Sc chains  \cite{Hui2005,Piskunov2016}. 
As one can see, the Sc--F--Sc BADF for the $2a \times 2a \times 2a$ supercell deviates significantly from the others being broader and giving the average value of the 
Sc--F--Sc angle by about $\sim 5^\circ$ smaller. This means that the simulations using the smallest supercell should fail in describing dynamical behaviour of ScF$_3$ and, in fact, result in the underestimated absolute values of the lattice parameter (Fig.~\ref{fig1}).

The effect of dynamic disorder can be illustrated using the comparison of the experimental and calculated Sc K-edge EXAFS spectra and their Fourier transforms (FTs) 
shown in Fig.~\ref{fig4} for the four supercell sizes at  $T$ = 600 K. Thermal disorder is responsible for the EXAFS amplitude damping at high-$k$ values \cite{Kuzmin2014IUCR} and, consequently, leads to a reduction of the peaks amplitude in the FTs.  The first peak at 1.6 \AA\ in FT corresponds purely to the first shell contribution due to 6 fluorine atoms (F$_1$).
The second peak at 3.5 \AA\ has complex origin due to the interference between the second (Sc$_2$) and third  (F$_3$) coordination shells plus the 
so-called multiple-scattering contributions
generated within the Sc--F--Sc atomic chains, which are sensitive to the  Sc--F$_1$--Sc$_2$ bond angle variation. The last peak at 5.4 \AA\ is mainly due to 
the fourth (Sc$_4$) and fifth  (F$_5$) coordination shells.
As one can see, the AIMD simulations reproduce well the experimental EXAFS data for all supercell sizes except the smallest one, for which 
the amplitude of all three peaks in FT is systematically smaller than in the experiment. This result correlates well with the behaviour of RDFs in Fig.~\ref{fig5}. 
It is interesting to note that in spite of the Sc--Sc$_2$ RDF is close for all supercells, the sensitivity of the FT peak at 3.5 \AA\ to  the  Sc--F$_1$--Sc$_2$ bond angle 
gives origin of its reduced amplitude for the $2a \times 2a \times 2a$ supercell.

\begin{figure}[t]
	\begin{center}
		\includegraphics[width=0.4\linewidth]{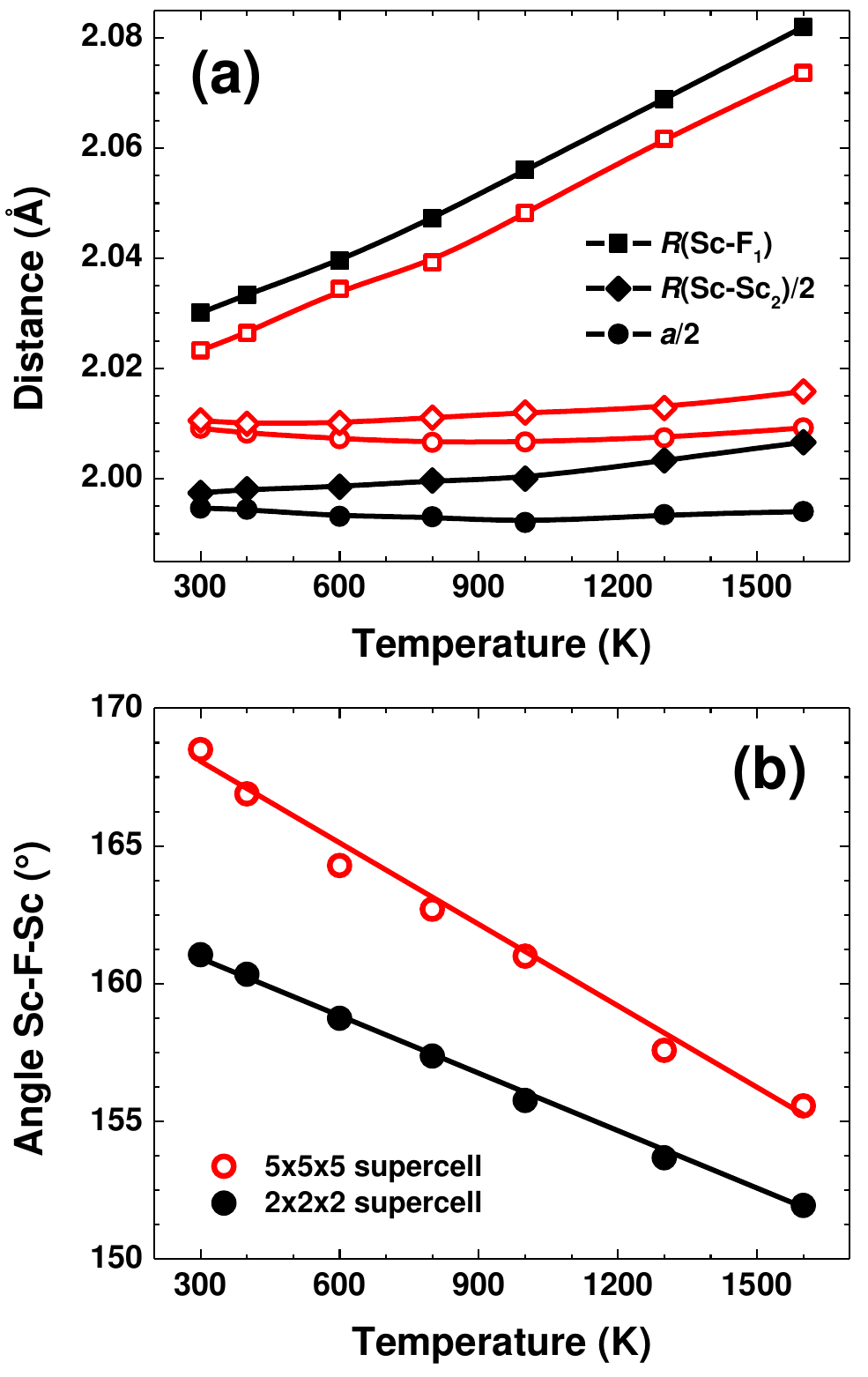}
	\end{center}
	\caption{(a) Temperature dependence of the lattice parameter $a$ (circles) and interatomic distances $R$(Sc--F$_1$) (squares) and $R$(Sc--Sc$_2$) (diamonds) in ScF$_3$ calculated by AIMD for the $2a \times 2a \times 2a$ (solid symbols) and $5a \times 5a \times 5a$ (open symbols) supercells. (b) Temperature dependence of the average bonding Sc--F-Sc angle in ScF$_3$ calculated by AIMD for the $2a \times 2a \times 2a$ (solid circles) and $5a \times 5a \times 5a$ (open circles) supercells. Lines are guides for the eye.}
	\label{fig7}
\end{figure}

The NTE mechanism in ScF$_3$ can be understood from our AIMD simulations
by considering the temperature variation of the lattice parameter $a$ and interatomic distances $R$(Sc--F$_1$) and $R$(Sc--Sc$_2$) (Fig.~\ref{fig7}(a)) and related variation of the average Sc--F--Sc bonding angle (Fig.~\ref{fig7}(b)).
For ease of comparison, the values of $a$  and $R$(Sc--Sc$_2$) are divided by two. Note that thermal vibrations of scandium atoms in the direction orthogonal to the crystallographic axes are responsible for the difference between the values of $a$ and $R$(Sc--Sc$_2$) \cite{Piskunov2016}. 

An increase of temperature affects strongly the Sc--F$_1$ bond, which elongates almost linearly by about 0.05 \AA\  in the temperature range of 300--1600 K. This trend is close in both supercell models and is in good agreement with the analysis of the experimental EXAFS data published in \cite{Hu2016,Piskunov2016} and the results of previous AIMD simulations from \cite{Lazar2015}. Such behaviour indicates that the ScF$_6$ octahedra do not behave as rigid units and expand significantly upon heating.  The behaviour of the Sc--F$_1$ bonds contrasts strongly with that of the Sc--Sc$_2$ interatomic distances suggesting much larger amplitude of thermal vibrations for fluorine atoms in the direction orthogonal to the Sc--F--Sc linkage. This fact is well evidenced 
by temperature variation of the average Sc--F--Sc bonding angle in Fig.~\ref{fig7}(b), which decreases upon heating meaning stronger rotations of ScF$_6$ octahedra.  

One should note that the $R$(Sc--Sc$_2$) next-nearest-neighbour distance behaves in a slightly different way for small and large supercells. 
It always expands (by about 0.02 \AA) for the $2a \times 2a \times 2a$ supercell in the temperature range of 300--1600 K, whereas it has a shallow minimum at 400 K and expands by only 0.01 \AA\ up to 1600 K for the $5a \times 5a \times 5a$ supercell.   Also the difference between $R$(Sc--F$_1$) and  $R$(Sc--Sc$_2$)/2 distances is larger for smaller supercell, indicating stronger
rotations of ScF$_6$ octahedra as is also evidenced in the Sc--F--Sc BADFs in Fig.~\ref{fig6}. 

Thus, while the NTE effect in ScF$_3$ is reproduced using all supercell models 
and can be explained by the interplay between expansion and rotation of ScF$_6$ octahedra,
the small supercell overestimates the contraction of the lattice and vibrational amplitudes of atoms. This fact is responsible for the worse agreement between the experimental and simulated Sc K-edge EXAFS spectra in Fig.~\ref{fig4} for the smallest $2a \times 2a \times 2a$ supercell.

\section{Conclusions}\label{conc}

The ab initio molecular dynamics (AIMD) simulations performed
within the isothermal–isobaric (NpT) ensemble as presented in this
study are able to reproduce the negative thermal expansion of ScF$_3$ up to 
$\sim 1000$ K and the positive expansion at higher temperatures in agreement with previous diffraction data \cite{JACS}. The origin of the NTE in ScF$_3$ is explained by the interplay between expansion and rotation of ScF$_6$ octahedra.

At the same time, the simulations based on the smallest supercell ($2a \times 2a \times 2a$) fail to describe thermal disorder accurately, 
leading to an overestimated broadening of the inter-octahedral Sc--F--Sc bond angle distribution and of the outer coordination shells (starting from the third) 
in the radial distribution functions of scandium. 

The results obtained by the AIMD simulations were validated using the MD-EXAFS approach based on the ab initio multiple-scattering theory. 
A comparison between the calculated and experimental Sc K-edge EXAFS spectra at $T$ = 600 K suggests that a supercell larger than $2a \times 2a \times 2a$ should 
be employed to obtain good agreement, and the best results are achieved for a supercell of at least $4a \times 4a \times 4a$. 
Thus, we demonstrated that the results of the AIMD simulations are sensitive 
to the size of the supercell, and the experimental EXAFS spectra can be used to distinguish between different theoretical models.

\section*{Acknowledgements}
The calculations were performed on the Paul Scherrer Institute cluster Merlin4, HPC resources of the Swiss National Supercomputing Centre (CSCS) in Lugano as well as at the Latvian SuperCluster (LASC). The authors gratefully acknowledge staff of the Swiss National Supercomputing Centre (CSCS) during the project ID s628 realization. 

The authors sincerely thank S. Ali, A. Kalinko, and F. Rocca for providing experimental EXAFS data, as well as M. Isupova, V. Kashcheyevs, and A. I. Popov for stimulating discussions. 
Financial support provided by project No. 1.1.1.2/VIAA/l/16/147 (1.1.1.2/16/I/001) under the activity ``Post-doctoral research aid'' realized at the Institute of Solid State Physics, University of Latvia is greatly acknowledged by D.B. 
A.K and J.P. would like to thank the support of the Latvian Council of Science project No. lzp-2018/2-0353. Authors have no conflict of interest to declare.

\section*{Data availability}
The raw/processed data required to reproduce these findings cannot be shared at this time as the data also forms part of an ongoing study. 

CRediT authorship contribution statement 

\textbf{D. Bocharov}: Conceptualization, Data curation, Formal analysis,
Funding acquisition, Investigation, Methodology, Project
administration, Resources, Validation, Visualization, Writing - original
draft, Writing - review \& editing. 
\textbf{M. Krack}: Conceptualization, Data
curation, Formal analysis, Investigation, Methodology, Resources,
Software, Supervision, Writing - review \& editing. 
\textbf{Yu. Rafalskij}: Data
curation, Formal analysis, Investigation, Software, Validation,
Visualization. 
\textbf{A. Kuzmin}: Conceptualization, Data curation, Formal
analysis, Investigation, Methodology, Resources, Software, Supervision,
Validation, Visualization, Writing - review \& editing. 
\textbf{J. Purans}: Conceptualization, Funding
acquisition, Investigation, Methodology, Project administration,
Writing - review \& editing.

\end{document}